\documentclass[preprint]{elsarticle}

\usepackage[utf8]{inputenc}
\usepackage[dvips,unicode]{hyperref}
\usepackage{subfig}
\usepackage{array}
\usepackage{url}
\hypersetup{
  pdftitle={HYMAD: Hybrid DTN-MANET Routing for Dense and Highly Dynamic Wireless Networks},
  pdfauthor={J. Whitbeck, V. Conan},
  pdfkeywords={Delay Tolerant Networks, Wireless Ad-Hoc Networks, Routing}
}

\begin{document}

\title{HYMAD: Hybrid DTN-MANET Routing for Dense and Highly Dynamic Wireless Networks\tnoteref{t1}}

\tnotetext[t1]{This article is an expanded version of work presented
  at the IEEE WoWMoM Workshop on Autonomic and Opportunistic
  Communications (AOC 2009)~\cite{whitbeck-hymad}. This paper presents
  all-new results based on a full implementation of HYMAD in the ONE DTN
  Simulator~\cite{ONE}, as well as a more detailed evaluation of HYMAD
  in terms of network dynamics, density, and overhead.}

\author[thales,lip6]{John~Whitbeck}
\ead{john.whitbeck@lip6.fr}

\author[thales]{Vania Conan}
\ead{vania.conan@fr.thalesgroup.com}

\address[thales]{Thal\`es Communications\\
  160 bd de Valmy - 92704 COLOMBES Cedex - France}
\address[lip6]{UPMC Paris Universitas \\
  Laboratoire d'Informatique de Paris 6 \\
  104 avenue du Pr\'esident Kennedy - 75016 Paris - France}

\begin{abstract}
  Delay/Disruption-Tolerant Network (DTN) protocols typically address
  sparse intermittently connected networks whereas Mobile Ad-hoc
  Network (MANET) protocols address the fairly stable and fully
  connected ones. But many intermediate situations may occur on
  mobility dynamics or radio link instability. In such cases, where
  the network frequently splits into evolving connected groups, none
  of the conventional routing paradigms (DTN or MANET) are fully
  satisfactory.  In this paper we propose HYMAD, a Hybrid DTN-MANET
  routing protocol which uses DTN between disjoint groups of nodes
  while using MANET routing within these groups. HYMAD is fully
  decentralized and only makes use of topological information
  exchanges between the nodes.  The strength of HYMAD lies in its
  ability to adapt to the changing connectivity patterns of the
  network. We evaluate the scheme in simulation by replaying synthetic
  and real life mobility traces which exhibit a broad range of
  connectivity dynamics. The results show that HYMAD introduces
  limited overhead and outperforms the multi-copy Spray-and-Wait DTN
  routing protocol it extends, both in terms of delivery ratio and
  delay.  This hybrid DTN-MANET approach offers a promising venue for
  the delivery of elastic data in mobile ad-hoc networks as it retains
  the resilience of a \textit{pure} DTN protocol while significantly
  improving performance.
\end{abstract}


\maketitle

\section{Introduction}
Wireless ad-hoc networking has emerged over the past decades at the
intersection of personal computing, cellular telephony, and the
Internet. It is best suited for use in situations where an
infrastructure is not available, or too costly to deploy. In a
wireless ad-hoc network, radio equipped devices (laptops, smart phones,
sensors, etc.) cooperate in a distributed manner to provide the
necessary network functionality in the absence of a fixed
infrastructure. Applications include for example emergency and rescue
operations, conference or campus settings, body area and personal
networks, or vehicular networks.

Node mobility has been acknowledged as one of the key challenges of
ad-hoc networking with direct impact on protocol performance. Early
models, such as Random Walk, Random Direction and Random Waypoint,
consider nodes moving on a free-space planar surface. More refined
characterization and modeling have been derived for pedestrian
applications~\cite{mcnamara,claveirole} and for vehicular
mobility~\cite{harri:models}.

The measurements and observations of mobility patterns, ranging from
students on a campus to pedestrians or taxi-cabs in a city, have
pushed the community to also consider extremely sparse networks, or
very dynamic mobility conditions, e.g. often observed in the vehicular
setting, where end-to-end paths between any pair of nodes may seldom
exist~\cite{Burgess:2006,crawdad}.

For such sparse or dynamic cases, the 
Delay/Disruption-Tolerant Network (DTN)~\cite{dtn_fall_sigcomm}
paradigm uses node mobility to its advantage while compromising on
message delivery delays~\cite{GrossglauserTse2002}. Message forwarding
decisions are made on a \textit{per-encounter} basis, for example by
using utility functions based on aggregating statistics on node
meeting probabilities~\cite{lindgren03,daly07,LER}. At any given
time, a node's vision of the network topology is limited to its
current neighbors, whereas conventional Mobile Ad-hoc Network (MANET)
routing schemes have either complete (in the proactive case) or at
least partial (in the reactive case) knowledge of the actual network
topology.

But even in the extreme case of sparse of highly mobile networks there
are situations where the network is sufficiently dense and well
connected to temporarily provide end-to-end connectivity between a
significant subset of its nodes.

These nodes may even form small islands of stability. Using MANET
principles within such islands can bring great improvements. Indeed,
it considerably increases each node's information of its local
topology, thus leading to better forwarding decisions.  When high
mobility rates and more generally high link instabilities reduce route
life-times and threaten network-wide end-to-end connectivity, a MANET
routing protocol can still succeed locally even if it fails globally.

In this paper we propose HYMAD, a Hybrid DTN-MANET routing protocol.
HYMAD combines techniques from both traditional ad-hoc routing and DTN
approaches. HYMAD periodically scans for network topology changes and
builds temporary disjoint groups of connected nodes. Intra-group
delivery is performed by a conventional ad-hoc routing protocol and
inter-group delivery by a DTN protocol.

HYMAD constantly adapts to the dynamics of the wireless ad-hoc network
using only topological information. As in traditional ad-hoc routing,
no extra information on geographical location or social community
membership is required. It does not rely on a priori knowledge of
connectivity patterns or inter-meeting times. This makes HYMAD
amenable to implementation in a DTN stack or ad-hoc routing
protocol~\cite{JOTT06}. In a dense network, HYMAD can function
similarly to a traditional MANET protocol. In the other extreme case
of very sparse connectivity (where only two nodes can be in contact at
any time) each node is a group on its own and HYMAD behaves like a
classical DTN routing protocol. In any other intermediate case its
hybrid nature takes over.

We implemented the HYMAD hybrid approach in the ONE DTN
simulator~\cite{ONE} with a self-stabilizing group
service~\cite{r_operators,DKP08} and the multi-copy Spray-and-Wait
protocol as the DTN routing scheme~\cite{spyro_sw}.

We evaluated the scheme by performing simulation runs both on
synthetic Random Waypoint mobility traces and on the real-life
Rollernet data set~\cite{tournoux08_rollernet}, an example of a highly
dynamic ad-hoc network.

In the next section, we further describe how this hybrid approach
positions itself compared to existing DTN and MANET approaches.  In
section~\ref{sec:hymad}, we describe the HYMAD routing protocol
principles. We explain how nodes can agree on forming disjoint groups
and how such groups rather than individual nodes can be used as the
basis for DTN routing. We then evaluate the scheme both on synthetic
mobility models and on a real data set, the Rollernet experiment, in
section~\ref{sec:results}.  The results show that HYMAD outperforms
the multi-copy Spray-and-Wait DTN routing protocol it extends, both in
terms of delivery ratio and delay.

\section{Routing in a mobile wireless network}
\label{sec:comparison}
\begin{figure}
  \centering
  \includegraphics{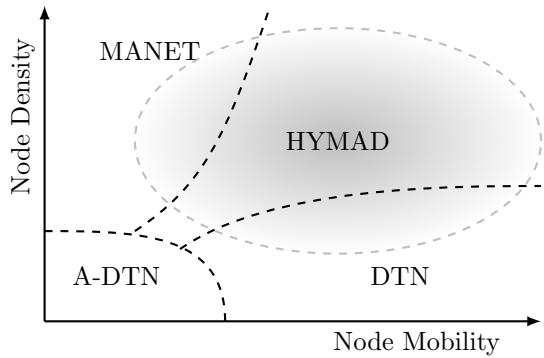}
  \caption{Mobility vs Density: when different paradigms apply}
  \label{fig:mob_density}
\end{figure}

Mobile wireless ad-hoc networks were first studied under the
assumptions of moderate node mobility and sufficient density to ensure
end-to-end connectivity. Both conditions are necessary for traditional
MANET approaches, be they proactive or reactive. 

Recently, there has been an effort to classify the various types of
mobile ad-hoc networks~\cite{AMKA08}. One can characterize the
relevant routing paradigms in mobile wireless networks along the two
main parameters of \emph{node density} and \emph{node mobility}. In
Fig.~\ref{fig:mob_density}, which maps the different routing
approaches on the bi-dimensional mobile wireless network space,
traditional MANET routing appears in the top left corner.

When the density of nodes diminishes, end-to-end connectivity can
disappear.  In such sparse networks, nodes have very few, if any,
neighbors within their transmission ranges. The topology eventually
splits into several non-communicating connected components.  This is
typically the realm of Delay Tolerant Networking which one can further
subdivide in two~\cite{BorrelAmmar07}: the Assisted DTNs (A-DTN), in
case of low mobility of nodes, or Unassisted DTNs (U-DTN) where
mobility is high. The latter corresponds to traditional DTN scenarios.

Routing in A-DTNs typically involves special mobile
nodes, known as message ferries or data mules, which relay the messages
between the separate connected components
~\cite{ZhaoFerries,Shah2003}. The packet-switching method of
MANETs is replaced with a store-and-forward approach.

When the mobility in sparse networks increases, mobile nodes begin to
meet others. This is the traditional DTN scenario, where nodes forward
one or more copies of a given message until it reaches its
destination. There are many strategies for optimizing the forwarding
decision. The most straightforward approaches, such as Epidemic or
Spray-and-Wait~\cite{spyro_sw} do not require nodes to acquire
information on the others' positions, movements or trajectories.  More
elaborate schemes involve a utility function where each node collects
direct and indirect knowledge of other nodes' meeting
probabilities. They require a certain learning period to aggregate
statistics before making good forwarding decisions. For example,
Lindgren et al.~\cite{lindgren03} use past encounters to predict the
probability of meeting a node again while Daly et al.~\cite{daly07}
use local estimates of betweeness and similarity.

In dense networks, conventional MANET protocols start to break down
under high mobility down even if the network is almost always fully
connected. Indeed the sheer instability of the links would result in a
deluge of topology updates in the proactive case and \textit{route
  error} and new \textit{route requests} messages in the reactive
case. DTNs protocols on the other hand can handle high mobility
regardless of the density of the network. However by narrowly focusing
on per-encounter events, they ignore a lot of available
information. For example, simply asking nodes to regularly broadcast
a list of their neighbors would give each node a picture of its
two-hop neighborhood even under high mobility. Repeat this once and
everyone knows their three hop neighborhood. A node may therefore
have a topology ``knowledge horizon'' which determines how far into
the real topology a node can ``see''. The more extreme the mobility,
the shorter the ``horizon''.

The Hybrid DTN-MANET approach that we advocate in this paper aims at
filling the gap for efficient routing in highly connected and highly
mobile networks, which have so far, to the best of our knowledge,
received little attention.  Hybrid DTN-MANET routing, like the HYMAD
protocol that we describe below, combines the resilience of DTNs with
the greater knowledge of local network topology provided by a MANET
protocol. It adapts naturally to the dynamics of the network and its
applicability spans a large spectrum of the mobile wireless network
space.

\section{The HYMAD protocol}
\label{sec:hymad}

\subsection{Overview}
\label{subsec:overview}

The core idea in HYMAD is to use whole groups of nodes instead of
individual nodes as the focus of a DTN protocol. The analogy is
detailed in Table~\ref{table:analogy}.

\begin{table}
  \centering
\caption{HYMAD groups vs. DTN nodes}
\begin{tabular}{m{2.5cm}m{5.5cm}}
DTN & HYMAD \\
\hline
Node & Group of nodes \\
\hline
A node has message $m$ & One node in the group has message $m$ and all other nodes in the group know that. \\
\hline
Two nodes meet & Two disjoint groups become connected. \\
\hline
\end{tabular}
  \label{table:analogy}
\end{table}

Each node $u$ regularly broadcasts, not necessarily with the same
frequency, two control messages:

\begin{enumerate}
\item An ``enhanced distance vector'' for all members of its group.
\item A list of the messages held by members of its group, with their respective
  destinations, custodians, and number of copies. 
\end{enumerate}

The first message enables the use of an intra-group distance vector
routing protocol. In this paper's implementation, the actual
transmitted distance vector is slightly modified with a node marking
mechanism for the distributed network partitioning algorithm
(Section~\ref{subsec:intra_group}). Furthermore, each node can be
tagged as a \emph{border node} (i.e. in contact with other groups). We
will refer to this list as the \emph{Group algorithm message}.

The second message is necessary for the inter-group routing
protocol. The message list allows a group to agree on what messages it
carries, and, for each message, which node (hereafter call the
message's custodian) specifically holds it. The border node tags give
every group member an estimation of the number of outgoing connections
to neighboring groups. We will refer to this list as the
\emph{Messages-in-group list}.

As in traditional distance vector algorithms, the number of iterative
broadcasts necessary for all members of a group to agree on this
information is equal to the diameter of the group.

HYMAD then uses a DTN protocol to transfer messages between
groups. The approach is generic and many existing DTN protocols could
be employed.  In this paper, Spray-and-wait~\cite{spyro_sw} is used to
forward messages between disjoint groups. As in Spray-and-Wait, the
source of a message will create a certain number of copies of it. In
HYMAD however, this source node is part of a group and copies of the
message will be distributed among the adjacent groups instead of
simply the nodes that the source encounters. If a group has more than
one copy, it will, in turn, distribute extra copies to its other
adjacent groups. If a group has just one copy it will wait until
encountering the destination's group to transfer it. Once inside the
destination's group, the intra-group routing protocol delivers the
message to the destination.

\subsection{Intra-group routing}
\label{subsec:intra_group}

\begin{figure}
  \centering
  \includegraphics{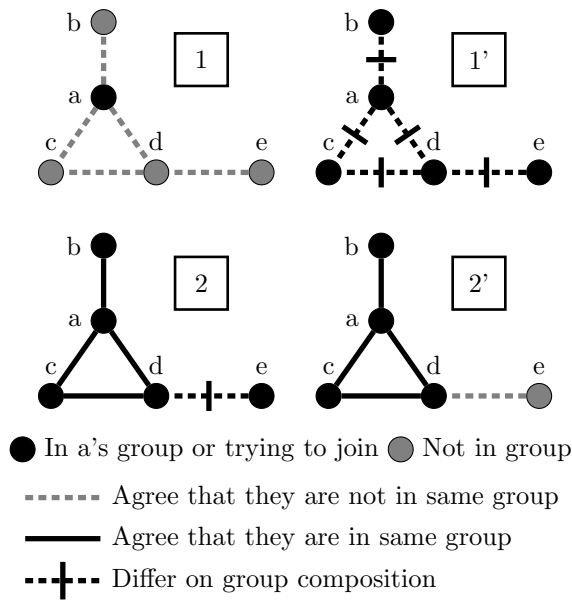}
  \caption{Self-stabilizing groups: convergence in two iterations}
  \label{fig:group_spread}
\end{figure}

In HYMAD, the intra-group routing is handled by a simple distance
vector algorithm. The only difference is that if a next hop is no
longer available, instead of discarding the message, the node takes
custody of it and adds it to its list of held messages in order to
advertise the fact to its group members.

That stills leaves open the question of how to partition the network
into several disjoint groups of nodes using a simple distributed
network partitioning algorithm. In our implementation, we chose to
consider \emph{diameter-constrained} groups. This is a particularly
relevant metric because it preserves node equality and
interchangeability. Indeed, another way of defining groups could have
involved considering all the nodes within two hops, for example, of a
given node. However this would mean that some nodes are somehow more
important than others. The diameter parameter on the other hand does
not refer to any particular node and only makes sense within a
connected subset of the nodes. It is therefore a natural candidate
for defining disjoint groups of nodes.

A group will accept new members as long as its diameter is less than a
predefined network-wide maximum diameter parameter ($D_{max}$). If a
group's diameter expands due to internal link failure, then some
members are excluded to satisfy the diameter constraint. Ducourthial
et al.~\cite{DKP08} propose a self-stabilizing, asynchronous
distributed algorithm that achieves this using an \mbox{$r-operator$}
on a slightly modified distance vector. This algorithm converges in
$O(D_{max})$ iterations. The proof of self-stabilization using
asynchronous message passing can be found in~\cite{r_operators}.

The main ideas behind group creation and modification are illustrated
in Fig.~\ref{fig:group_spread} for a maximum diameter $D_{max}=2$. In the
first iteration, node $a$ begins by broadcasting the distance vector
$(a:0)$. Nodes $b$, $c$ and $d$ decide they want to join the group and
broadcast $(b:0,a:1)$, $(c:0,a:1)$, and $(d:0,a:1)$, respectively. After
receiving the broadcast from $d$, node $e$ also decides that it wants
to join the group and broadcasts $(e:0,d:1,a:2)$ (or
$(e:0,d:1,c:1,a:2)$ if $c$ spoke before $d$). In the second iteration,
$a$ now broadcasts $(a:0,b:1,c:1,d:1)$, $d$ realizes that the distance
between $b$ and $e$ is greater than $D_{max}$ and therefore chooses to
exclude $e$ from the group and broadcasts $(d:0,a:1,c:1,b:2)$. Finally
$e$ understands that it is not part of the group. After two
iterations, the group has stabilized on $a,b,c,d$. Now lets suppose
that at a later time the link between $a$ and $c$ goes down. Node $c$
now only receives the broadcasted distance vector $(d:0,a:1,b:2)$ from
$d$. It then understands that it is no longer part of the group. As is
obvious from this example, a given topology can result in very
different groups depending on the order in which the nodes speak.

Some looping problems can arise if two nodes simultaneously try to
join a group that can only accept one of them. In order to avoid this,
Ducourthial et al. suggest a priority mechanism, though refrain from
actually detailing one~\cite{DKP08}. In order to work, the priority
function has to define a total order on the nodes. An interesting goal
would be to ensure that a stable group cannot be split up when a new
node, which happens to have priority over some group members, comes
within range. We achieve this by defining each node's priority as
either the time when it joined its current group or the current time
if it is not in a group. We add to this a small value unique to each
node that guarantees that a total ordering is defined at all
times. These priority scores are attached to the \emph{Group algorithm
  message}.

As in some MANET protocols, this algorithm is used in a proactive
fashion where each node periodically runs the algorithm and broadcasts
its modified distance vector. Group composition therefore changes in
reaction to topology changes rather than routing needs. Unlike most
MANET protocols, this information is never propagated beyond a group
and its immediate neighbors. The size of the \emph{Messages-in-group
  lists} is proportional to the number of nodes in a group, while the
convergence time for all group information is in $O(D_{max})$.

\subsection{Inter-group routing}
\label{subsec:inter_group}

\textit{Border nodes} take care of most of the inter-group DTN
routing. Indeed, the periodic broadcast protocol described in
\ref{subsec:intra_group} puts them in the unique position of knowing both the
composition of two adjacent groups as well as the messages these
hold. \textit{Border nodes} may request a message's custodian to
transfer one or more copies to it.

When a \textit{border node} learns that its group has acquired copies
of a message that a neighboring group does not possess, it has the
following options:
\begin{itemize}
\item If the message's destination is in the neighboring group,
  request the message from its custodian and pass it on.
\item If its group has more than one copy of the message, request
  $max\left(1,\left\lceil \frac{n_c}{n_b+1} \right\rceil \right)$ copies from its
  custodian and pass them on.  ($n_c$ is the number of copies and
  $n_b$ the current number of border nodes in the group). The idea is
  to fairly spread a group's copies among its adjacent groups.
\item Otherwise do nothing
\end{itemize}

Conversely, when a \textit{border node} receives copies of a new
message from an adjacent group it can either:
\begin{itemize}
\item If the destination is in its group, forward the message to
  it using the inter-group routing protocol.
\item Otherwise, randomly select a group member to be the custodian
  for the copies. This is done to spread the burden over members of a
  group.
\end{itemize}

With this in place, when a node wants to send a message, it simply
adds it to its own list of messages. Through the intra-routing protocol,
in $O(D_{max})$ time, the group's \textit{border nodes} will become
aware of the new message and request copies to forward it on to the
adjacent groups.

In order to achieve this, two types of control messages are needed: a
\emph{Copy Request} message to ask a node in a group to forward a
copy, and a \emph{Reduce Number of Copies} message to ask a node in a
group to reduce the number of copies it currently holds. The latter
message generates a response indicating how many copies many the node
actually managed to remove. The importance of these two control
messages is detailed in section~\ref{subsubsec:copies} below, but it
is important to stress that these messages are very small and their
overhead is negligible compared to the overhead of the \emph{Group
  algorithm message} and \emph{Messages-in-group list} that the nodes
periodically broadcast.

\subsection{Example}
\label{subsec:example}

\begin{figure}
  \centering
  \includegraphics{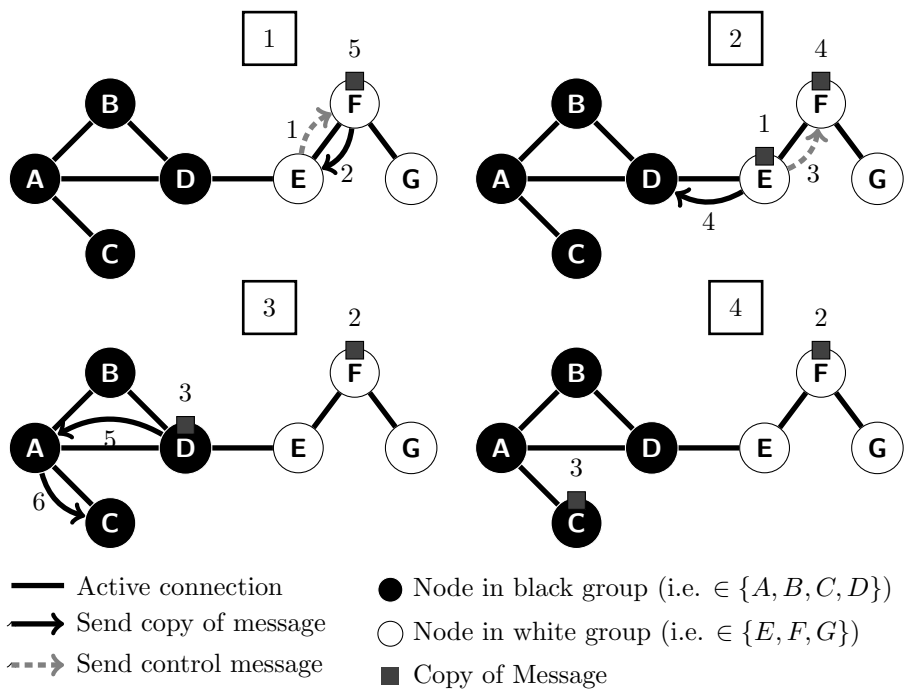}
  \caption{Example of message forwarding between two HYMAD
    Groups. (Full explanation in section~\ref{subsec:example})}
  \label{fig:hymad_example}
\end{figure}

Fig.~\ref{fig:hymad_example} details an example of HYMAD message
forwarding. In this case, the HYMAD groups have stabilized into two disjoint
groups (of diameter 2), a black one containing nodes A, B, C, and D,
and a white one containing nodes F, E, and G. Node F creates 5 copies
of a message for node C. It then adds the message to its list of held
messages. Upon receiving F's broadcasted list, node E becomes aware of
the new message. Being a \emph{border node}, E periodically receives
node D's distance vector and therefore knows that the message's
destination is a member of the black group. E therefore requests (1) a
copy from F, who forwards it on (2). Before transferring the message
on to D, F believes (correctly) that it is white group's only border
node and thus determines that it must send 3 copies to the black
group. E then asks F to reduce its number of copies by 2, to which F
answers positively (3). Now that E has its 3 copies, it forwards them
on to D (4). D receives these copies and realizes that they are for
C. Through its intra-group routing protocol, D knows that the shortest
path to C runs through A. It therefore transmits the message to A (5),
which then forwards it to C (6) for a successful delivery.

Had C not been the final destination for the message, steps 1-4 would
still have occurred, because E would have compared the message lists
from both the black and white groups and determined that the black
group did not have a copy of the message. Upon receiving the message,
node D would have chosen a random custodian for the message within its
own group and then forwarded the 3 copies to it.

\subsection{Discussion}
\label{subsec:discussion}

\subsubsection{Reacting to topology changes}
In the previous example, the connectivity conditions were idyllic. But
what would have happened if the link between D and E had come down during the
message transfer? The white group would then have had two custodians of
the message, with different numbers of copies. What if the link
between A and D had broken, thus extending black group's diameter to
3? Either D or C would have been excluded from it.

There are a number of situations, such as when two groups merge, where
copies of a given message can be distributed among several group
members. This is not a problem because everyone in the group will be
aware of this situation through the \emph{Messages-in-group list}
broadcasts. When a border node wishes to send some copies to a
neighboring group, it will consider the total number of copies in the
group, request a copy from one custodian, determine how many copies to
send, and then iteratively ask each custodian to provide enough copies
through \emph{Reduce Number of Copies} messages. Over time, the group
will return to the stable situation of a unique custodian with the
last remaining copy of the message for the group.

An internal link failure may increase a group's diameter, thus causing
it to split into several separate sub-groups. In such a situation,
each sub-group only has a fraction of the messages of the original
group. Fortunately this is not really a problem. Firstly, the
intra-group protocol detailed in section~\ref{subsec:intra_group}
ensures that nodes will update their message lists accordingly when
removing nodes from their group member lists, thereby preventing a
sub-group from advertising messages it does not have or any other such
incoherences. Secondly, certain subgroups may still be connected to
each other. If either sub-group has more than one copy of some
messages, these will be copied over the other sub-group. In any case,
HYMAD recovers gracefully from group splits.

\subsubsection{Managing the number of copies in a group}
\label{subsubsec:copies}

Because a message can have several custodians within a group, the
\emph{Reduce Number of Copies} signals are necessary to avoid
unnecessary retransmissions of the message. Indeed, it would be
wasteful for a border node to request copies from each custodian. It
only needs to get one complete copy, but has to make sure that the
extra custodians reduce their number of copies accordingly so as not
to create extra copies in the network. A subtle situation appears when
several border nodes \emph{concurrently} demand copies of a same
message. How many should each get? We solve this issue by determining
the number of copies to send at the last possible moment, right before
forwarding the message to the next group. There are therefore three
steps: (i) first send a \emph{Copy Request} message in order to
receive a complete copy, (ii) then estimate the number of copies held
by the group and the number of border nodes to calculate the number
$n$ of copies to forward on to the neighboring group, (iii) and finally
use of the \emph{Reduce Number of Copies} messages to ask the
message's custodians to collectively reduce their number of copies by
$n$. However, the custodians may only be able to reduce their number
of copies by $m<n$ if another border node has requested copies in the
meantime. In that case, $m+1$ copies are forwarded on to the
neighboring group. This mechanism ensures that the number of copies of
a message in the network remains constant at all times.

\subsubsection{The cost of clustering}

Group partitioning, or clustering, in a mobile ad-hoc network is
considered a costly and difficult problem. In many approaches, the
overhead increases with the network dynamics, a single change in
topology can lead to a complete re-clustering of the network, and
regular routing cannot function until the the clusters have
stabilized~\cite{ClusteringSurvey}.

However none of these problems arise in our approach. Indeed, since
the \emph{Group algorithm messages} are broadcast with a predefined
network-wide period, the speed of topology changes has no effect how
often these messages are broadcast. On the contrary, high dynamics
leads to unstable groups, which means many singleton groups, shorter
\emph{Group algorithm messages}, and ultimately lower overhead. For a
more thorough discussion of overhead, refer to
Section~\ref{subsec:overhead}. Furthermore, changes in one group's
composition do not propagate to the entire network. If one or more
nodes leave a given group, for example when a link goes down, they can
only join neighboring groups if they do not increase those groups'
diameter beyond $D_{max}$. Otherwise they are simply rejected and
remain alone. All group composition changes remain local. Finally,
HYMAD does not require nodes to agree on group composition. For
example, if a message is being sent through the intra-group routing
protocol and an intermediary node realises that the destination is not
in its group, he will simply stop the routing and add the message to his
own list of held messages. In any case, HYMAD recovers gracefully from
any inconsistencies.

\subsubsection{Choosing the diameter parameter}

Choosing a diameter parameter for the group self-stabilization
algorithm involves a trade-off. On the one hand, increasing it will
expand each node's individual ``knowledge horizon'' of the actual
network topology. Fewer copies will cover a larger portion of the
network, which will naturally lead to faster delivery. On the other
hand, this comes at the cost of increasing the convergence time of the
group service. It may also slightly increase the group service's overhead
(Section~\ref{subsec:overhead}). Ideally, the convergence speed should
be considerably faster than the speed of topology changes. In a sense,
extreme mobility may fundamentally limit a node's possible knowledge
of the network's topology.

If one is willing to incur the extra cost, the diameter can be set to
encompass the entire network. In such a situation, HYMAD resembles a
resilient MANET routing protocol using store-and-forward for message
transfers. Furthermore, in many mobile wireless scenarios, there are
underlying social dynamics at work which can sometimes drive nodes to
gather into loose communities. $D_{max}$ should be chosen so as to
allow the expected number of members per social group to neatly fit
into one self-stabilizing group.

\section{Simulation Results}
\label{sec:results}

\subsection{Methodology}
We implemented HYMAD in the ONE DTN simulator~\cite{ONE} in a
completely distributed fashion. The ONE simulator allows for testing
many different parameters including buffer sizes, transmission speeds
and ranges, message sizes, etc. It also implements a simple
interference model which forbids a node from communicating (either
receiving or sending) with two neighbors at the same time. In our
default scenario (Figs.~\ref{fig:100k}, \ref{fig:intermediate},
\ref{fig:5copies}), we considered 100MB buffers, 10kB message sizes,
100kB/s transmission speeds and infinite TTLs. The message generator
randomly creates a new message every second. In such a setup, buffer
size and channel congestion would not be issues. However, lower
transmission speeds could cause channel congestion. This, and how it
interacts with the HYMAD overhead, will be examined in the next
section.

Two types of simulations were conducted: the first uses the Random
Waypoint mobility model and the second replays the Rollernet
connectivity trace. For the Random Waypoint simulations, we kept the
default ONE movement parameters (min speed: $0.5m/s$, max speed:
$1.5m/s$, wait time: $2s$) and experimented with different numbers of
nodes and transmission ranges to study different densities, in order
to explore the two dimensional mobile wireless network space of
Fig.~\ref{fig:mob_density}. We also made the link capacity vary in
order to observe the network's behavior under congestion.

The results in this section are all averages obtained from 20 runs of
a given scenario. A set of 20 different seeds were used in each run to
initialize the message generators and the node mobility. Each
simulation ran an extra 1000 seconds after the last message was
created to give all sent messages time to arrive or expire. We
calculated 95\% confidence intervals for each measurement. These tend
to be quite tight (typically $\pm 0.02$) and are not reported on this
section's plots for clarity's sake.

We always compare HYMAD to both Epidemic and the regular
Spray-and-Wait it extends. In the default scenario, there is little
contention and Epidemic therefore provides an upper bound on
achievable performance in terms of both delay and delivery ratio. This
is not always true, particularly when the links are saturated
(Fig.~\ref{fig:1k}). Spray-and-Wait provides a DTN state-of-the-art
comparison.

Unless otherwise specified, we used $D_{max}=2$ for most simulations
because it ensures a very fast convergence rate and seemed to
accurately reflect the size of separate connected components (small
groups of friends for example), particularly during the accelerating
phases of Rollernet (Section~\ref{subsubsec:rollernet}). Because
of this, greater values of $D_{max}$ yield only a small improvement of
the delivery ratio. Interestingly, greater values of $D_{max}$ do not
noticeably increase the overhead, as discussed in the next section.

\subsection{HYMAD Overhead}
\label{subsec:overhead}

\begin{table}
  \centering
\caption{Overhead of HYMAD Control Messages}
\begin{tabular}{ll}
Message & Size (bits) \\
\hline
\emph{Copy Request} & $128$ \\
\hline
\emph{Reduce Number of Copies} & $128$ \\
\hline
\emph{Group algorithm messages} & $128 \times N_{group\;members}$ \\
\hline
\emph{Messages-in-group list} & $128 \times N_{messages\;in\;group}$ \\
\hline
\end{tabular}
  \label{table:overhead}
\end{table}

HYMAD makes use of several types of control messages. Two are
periodic, the \textit{Group algorithm message} and
\emph{Messages-in-group list} (Section~\ref{subsec:overview}), while
the other two, \emph{Copy Request} and \emph{Reduce Number of Copies}, 
are spontaneous (Section~\ref{subsec:inter_group}). In order to
estimate how large these control messages are, we assume that all node
ids, messages ids, and priority scores can fit on 32 bits. For
example, in ONE, a \emph{Copy Request} message only needs to contain
the id of the requesting node and the id of the requested message,
hence 64 bits, which was ``rounded up'' to 128 bits. On the other
hand, \emph{Messages-in-group list} varies with the
number of messages. The values and formulas used in the rest of this
section are in Table~\ref{table:overhead}. In this paper, in the
context of the ONE simulator, we consider that neither Epidemic nor
Spray-and-Wait incur any control overhead.

\begin{figure}[t]
  \centering
  \subfloat[2s \label{fig:overhead_2}]{\includegraphics{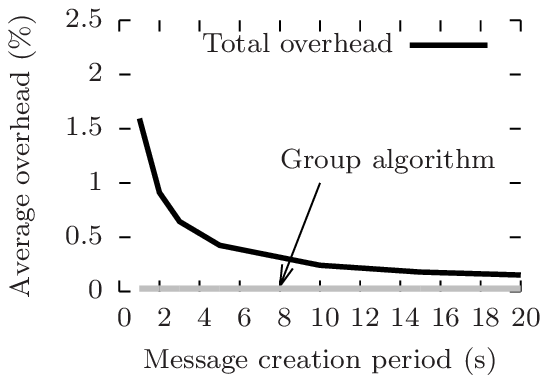}} \quad
  \subfloat[100ms \label{fig:overhead_01}]{\includegraphics{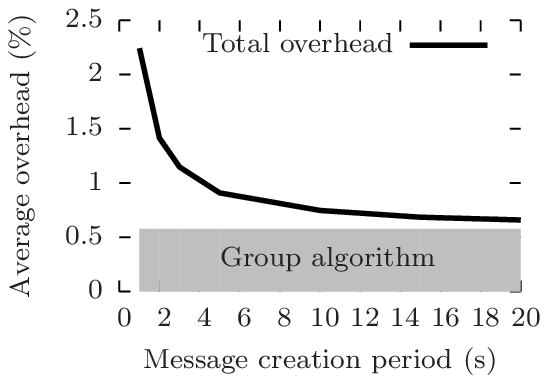}}
  \caption{HYMAD average link overhead vs Time period of appearance of new
    messages in the network. The \emph{Messages-in-group list} is
    broadcast every 2 seconds, whereas the \emph{Group algorithm
      messages} is broadcast every 2s in Fig.~\ref{fig:overhead_2} and
    every 100ms in Fig.~\ref{fig:overhead_01}.}
  \label{fig:overhead}
\end{figure}

Fig.~\ref{fig:overhead} plots the average link overhead depending on
how fast new messages are appearing in the network. The overhead
attributable to the \emph{Group algorithm messages} is invariant and
represents a lower asymptote for the total overhead. Indeed, even in
the absence of any messages, the group partitioning algorithm would
still be running. In Fig.~\ref{fig:overhead_2}, where both the
\emph{Group algorithm messages} and the \emph{Messages-in-group lists}
are being broadcast every 2 seconds, the group algorithm overhead is
negligible. In Fig.~\ref{fig:overhead_01}, the \emph{Group algorithm
  message}'s broadcast frequency was increased. Even when broadcast
every $100ms$ the group partitioning algorithm's overhead remains very
reasonable. In the rest of this paper, $100ms$ will be the default
broadcasting period for the group partitioning algorithm. In both
cases, the total overhead decreases non-linearly with the
\emph{Messages-in-group list} broadcasting period. Indeed, at greater
frequencies, messages appear faster than they can be delivered. This
increases the length of the \emph{Messages-in-group lists}, which
worsens the overhead, which in turn diminishes the link capacity
available to deliver the numerous messages.

The same measurements where made by varying the HYMAD diameter
parameter ($D_{max}$)instead of the broadcast frequency. Surprisingly, the
diameter parameter has very little impact on the total
overhead. Intuitively, a greater group diameter parameter should lead
to larger groups, and hence larger \emph{Group algorithm messages} and
\emph{Messages-in-group lists}, and ultimately an overhead
increase. However, closer examination shows that this is not the
case. Firstly, the topology limits how large groups can grow. For
example, increasing your diameter to 4 if most connected components
have a diameter of 2 or 3 will have little effect on
overhead. Furthermore, even if the effect were noticeable (e.g. in a
very dense scenario), the total overhead of the \emph{Group algorithm
  messages}, as explained previously, would still be
negligible. Secondly, lets consider a connected component that would
fit either two 2-diameter groups of diameter or a single 4-diameter
group. In the latter case, we expect the size of the
\emph{Messages-in-group list} to be proportional to the total number
of messages in the 4-diameter group (i.e. the entire connected
component), whereas in the former case, we would expect those control
messages to be proportional to the number of messages initially held
in each 2-diameter group, i.e. roughly half of those in the 4-diameter
group. However the HYMAD protocol ensures that copies will be passed
on between the two 2-diameter adjacent groups. Therefore, each 2-diameter group will
eventually hold the same number of messages as the 4-diameter group,
and their \emph{Messages-in-group lists} will be exactly as
long. Therefore, it follows that the diameter should have a
negligible effect on total overhead.

\begin{figure}[t]
  \centering
  \subfloat[100 kB/s \label{fig:100k}]{\includegraphics{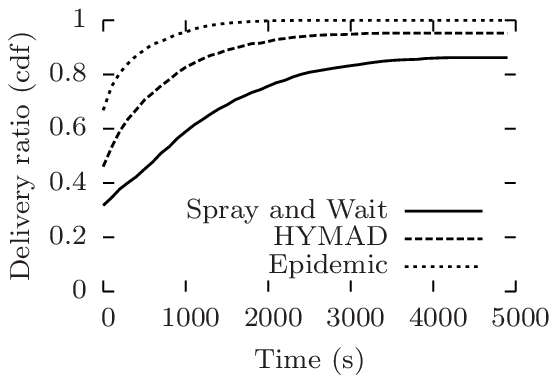}} \quad
  \subfloat[10 kB/s \label{fig:10k}]{\includegraphics{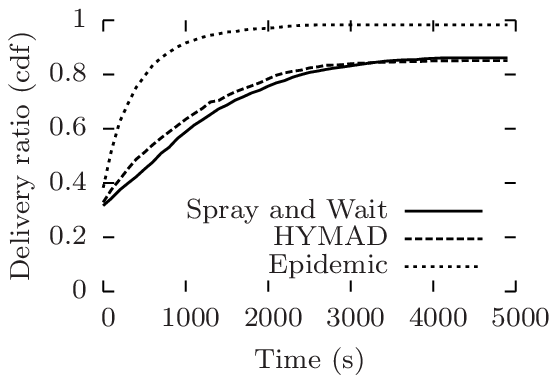}} \\
  \subfloat[1 kB/s \label{fig:1k}]{\includegraphics{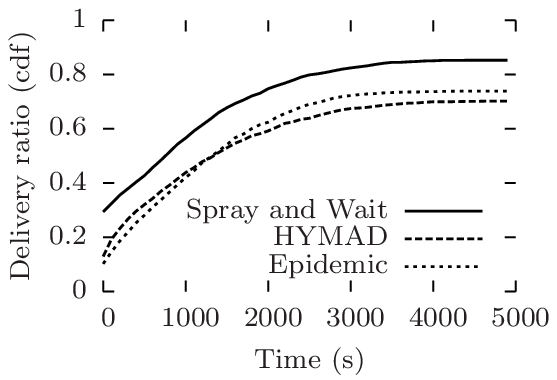}}
  \caption{Cumulative delivery ratio using Random Waypoint with 30
    nodes and varying transmission speeds.}
  \label{fig:rwp_capacity}
\end{figure}

Fig.~\ref{fig:rwp_capacity} compares the performance of HYMAD,
Epidemic and Spray-and-wait, with increasingly slow transmission
speeds. Fig.~\ref{fig:100k} is our default scenario. When the
transmission speed if divided by 10 (Fig.~\ref{fig:10k}), HYMAD's
overhead and occasionally useless retransmissions (e.g. when a group
splits) really start to degrade performance. When the transmission
speed drops to 1 kB/s, Epidemic routing's performance collapses as it
saturates the links. In these very challenged conditions,
Spray-and-wait actually performs the best.

All the results for HYMAD in this paper were obtained by first
measuring the average per-link overhead and then subtracting that
amount from the available link capacity in the ONE simulator.

\subsection{Influence of network connectivity}

\begin{figure}[t]
  \centering
  \subfloat[Sparse ($\bar{d}=0.6 \pm 0.1$, $R=400m$) \label{fig:sparse}]{\includegraphics{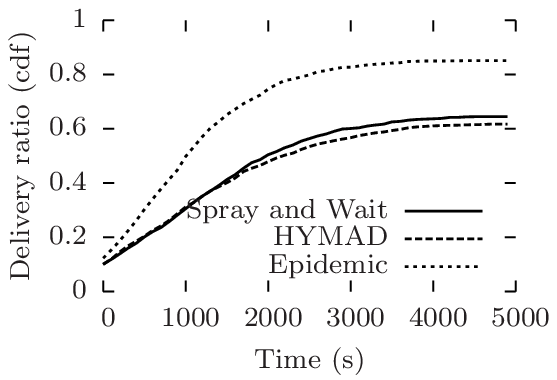}} \quad
  \subfloat[Medium ($\bar{d}=1.6 \pm 0.4$, $R=700m$) \label{fig:intermediate}]{\includegraphics{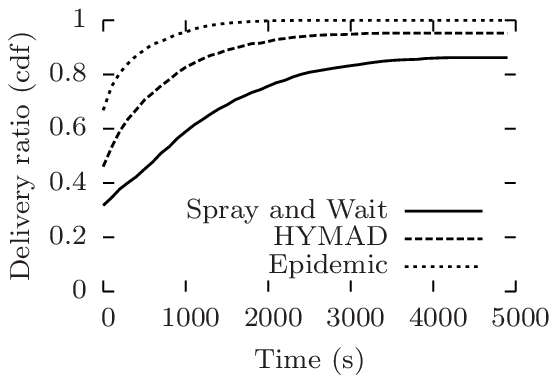}} \\
  \subfloat[Dense ($\bar{d}=6.7 \pm 0.6$, $R=1km$) \label{fig:dense}]{\includegraphics{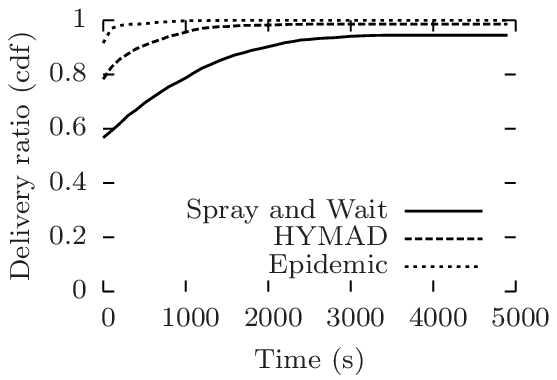}}
  \caption{Cumulative delivery ratio using Random Waypoint with 30
    nodes and varying transmission ranges (and thus varying node
    densities). The average node degrees ($\bar{d}$) and transmission
    ranges ($R$) are indicated in parentheses}
  \label{fig:rwp}
\end{figure}

Fig.~\ref{fig:rwp} compares HYMAD's delivery ratio to that of both
Epidemic and Spray-and-Wait for different network densities. We
considered 30 nodes all moving according to the Random Wapyoint
model. The only parameter that changes between these 3 scenarios is
the transmission range and hence the network density.  

When the network is very sparse (Fig.~\ref{fig:sparse}), nodes only
occasionally encounter others, and rarely more than one at a
time. Under such conditions, it is a good idea to distribute copies to
all the nodes that one encounters, as in Spray-and-Wait. On the other
hand, two HYMAD nodes may form a temporary two-node group in which
neither node transmits any of its messages to the other. In the sparse
case, HYMAD's group mechanisms actually degrade the delivery ratio,
compared to a more straightforward protocol like Spray-and-Wait.

When the network's density increases, HYMAD's performance overtakes
Spray-and-Wait's (Fig.~\ref{fig:intermediate}, the default scenario)
to the point of nearly matching the performance of Epidemic for dense
networks (Fig~\ref{fig:dense}). This is because HYMAD's knowledge of
its local network topology expands with the node density. Indeed,
there is a relation between the average group size, the number of
copies and the number of copies necessary to cover the network quickly
and efficiently. For example, in Fig.~\ref{fig:dense}, end-to-end
connectivity almost always exists, and we expect $D_{max}=2$ groups to
typically contain 4 or 5 nodes out of the total 30. Since the
algorithm tends to leave a single copy per group, there is a good
chance that the five copies will quickly spread to all the groups, one
of which is likely to contain the destination. This allows a message
to rapidly zero in on its destination and accounts for the
Epidemic-like performance in the dense scenario.

\subsection{Influence of the number of copies}

\begin{figure}[t]
  \centering
  \subfloat[5 copies \label{fig:5copies}]{\includegraphics{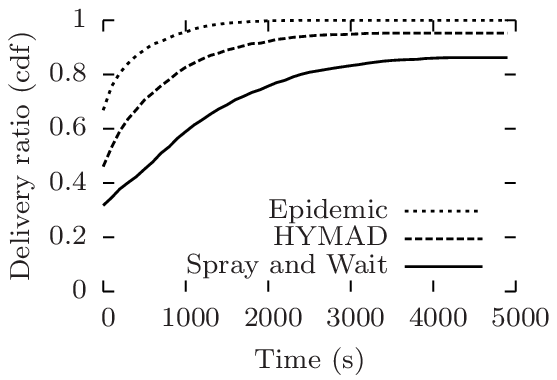}} \quad
  \subfloat[15 copies \label{fig:12copies}]{\includegraphics{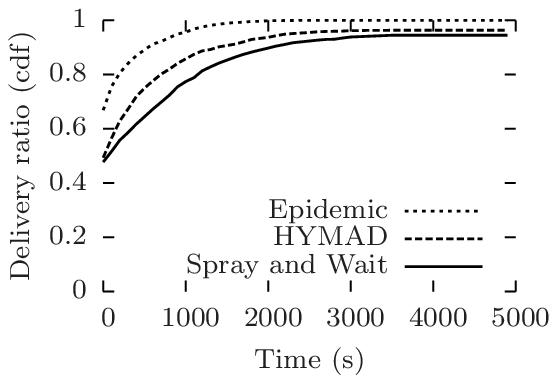}} \quad
  \caption{Impact of the number of copies on delivery ratio}
  \label{fig:rwp_copies}
\end{figure}

Fig.~\ref{fig:rwp_copies} compares the delivery ratio obtained by
HYMAD and Spray-and-Wait. Fig.~\ref{fig:5copies} again corresponds to our default scenario.
Predictably, the
performance of both Epidemic and Spray-and-Wait increases with the
number of copies.

\subsection{Results on the Rollernet trace}

\subsubsection{Rollernet}
\label{subsubsec:rollernet}

\begin{figure}[t]
  \centering
  \subfloat[Average node degree \label{fig:avg_node_degree}]{\includegraphics{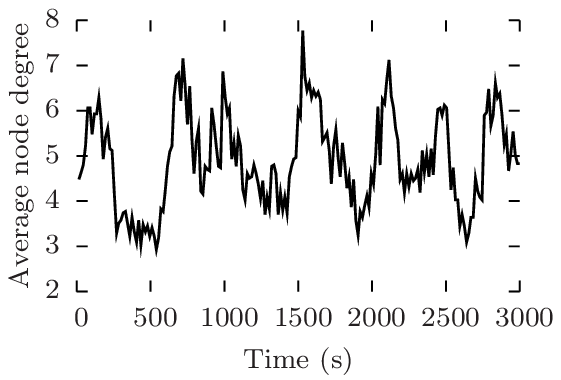}} \quad
  \subfloat[Number of connected components (ccs) \label{fig:num_ccs}]{\includegraphics{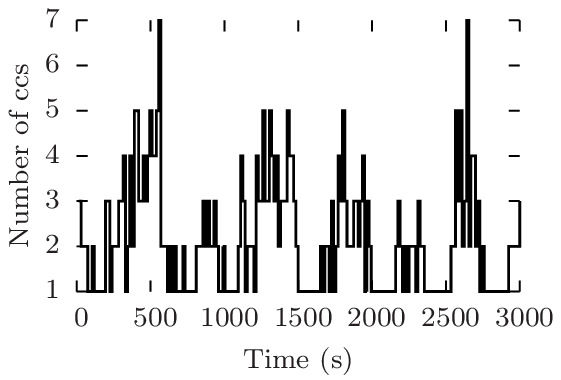}}
  \caption{The accordion effect in Rollernet}
  \label{fig:accordeon}
\end{figure}

In this section, we evaluate HYMAD's performance on
Rollernet~\cite{tournoux08_rollernet}, a highly connected and
extremely mobile connectivity trace. The Rollernet experiment involved
equipping 62 participants of the regular Sunday afternoon
rollerblading tour through Paris with contact loggers (Intel iMotes),
whose sampling period was set to 15 seconds. In order to witness
different behavior profiles, the 62 bluetooth loggers were distributed
among groups of friends, members of rollerblading associations, and
staff operators. In particular, one member of the staff was instructed
to remain behind the tour at all times while another stayed in front for the
entire duration of the experiment. This allows us to get a rough sense
of the relative geographic position of the participants by looking at
the connectivity graph. A snapshot of the connectivity graph can be
seen in Fig.~\ref{fig:diffusion} and an animation is available
online~\cite{rollernet_youtube}.

The Rollernet trace is ideal for evaluating HYMAD. Indeed it exhibits
the following characteristics:
\begin{itemize}
\item \emph{High density:} Contrary to many DTN traces, Rollernet is
  \emph{not} sparse. A look at Fig.~\ref{fig:avg_node_degree}, shows that
  the average node degree of the connectivity graph oscillates between
  2.9 and 7.8. The average for the whole tour is 4.8.
\item \emph{High mobility:} Everyone eventually meets everyone
  else. On average, each of the 62 nodes meets 56 others during the
  course of the tour. Additionally the topology evolves extremely
  quickly. The average lifetime of a given link is 26 seconds. The
  average lifetime of a shortest path between two nodes is 15.5
  seconds. Considering that the sampling period is 15 seconds, it
  follows that links are highly unstable and valid routes transient.
\item \emph{Accordion Effect:} This is an interesting consequence of
  the rollerblading context. The tour alternates between acceleration
  and deceleration phases in which the network topology respectively
  expands, leading to several separate connected component, and
  contracts, leading to a single connected
  component. Fig.~\ref{fig:num_ccs} shows that the number of connected
  components varies between 1 and 7 (17 if counting isolated
  nodes). In fact, Figures \ref{fig:avg_node_degree} and \ref{fig:num_ccs} have
  roughly alternating phases.
\end{itemize}

Although the accordion effect may be fairly specific to the Rollernet
context, the high density and mobility are not, and the insights
obtained by replaying this contact trace have a broad application to
DTNs ranging from networks of hand-held mobile devices to vehicular
ad-hoc networks.

\subsubsection{Group stability}

\begin{figure}[t]
  \centering
  \includegraphics{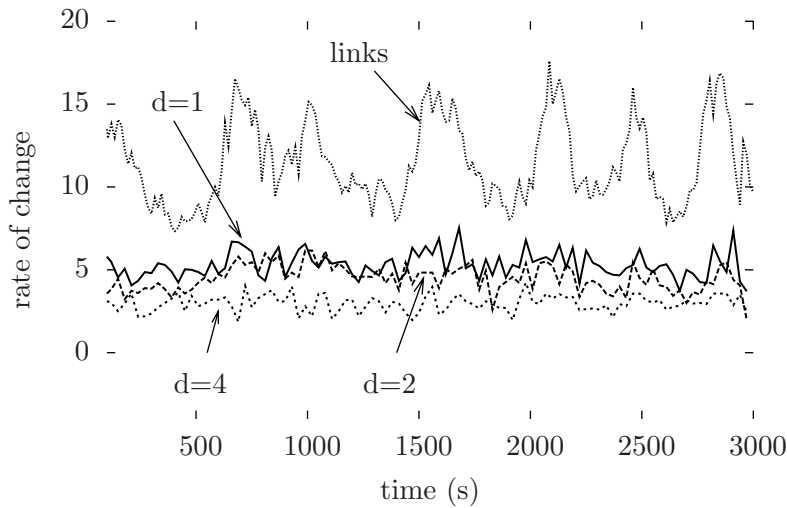}
  \caption{Link and group stability in Rollernet: number of links that
    come up or down per second, and number of nodes per second that either join
    or leave a group for varying values of $D_{max}$.}
  \label{fig:group_stab}
\end{figure}

Fig.~\ref{fig:group_stab} compares the rate at which links come up or
down in Rollernet with the number of nodes per second that either
join or leave a group for $D_{max} \in \{1,2,4\}$. 

Rate of change in group composition is very steady whereas the rate of
link changes follows the Rollernet accordion effect. This supports the
idea that small communities like groups of friends tend to stick
together during the tour and that link failures do not necessarily
mean that two nodes have clearly moved away from each other. The HYMAD
group structure appears to smooth the accordion effect.

Increasing $D_{max}$ reduces the rate at which nodes join or leave
groups. Indeed, even with $D_{max}=4$, the diameter of many groups
will be less than $4$.  A single link failure is much less likely to
increase such groups' diameter beyond $D_{max}$ and cause nodes to
leave. In this case, increasing $D_{max}$ makes groups more stable.

\subsubsection{Performance}

\begin{figure}[t]
  \centering
  \includegraphics{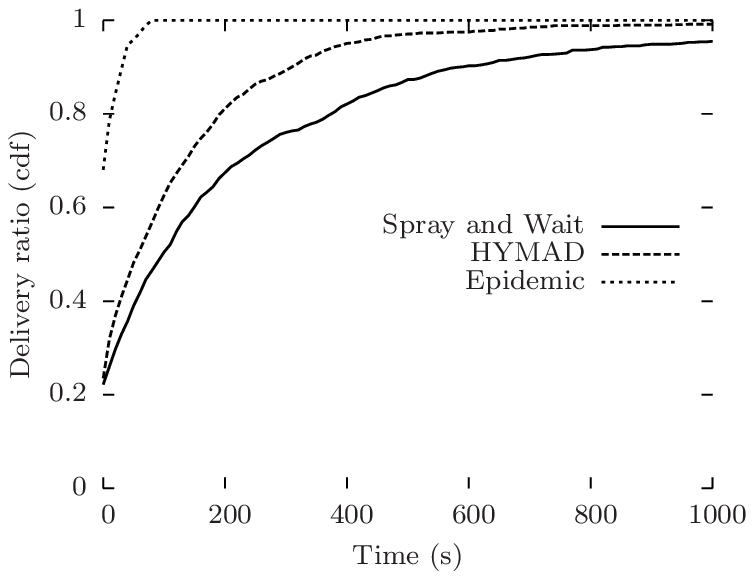}
  \caption{Rollernet: Comparison of delivery probabilities (5 copies).}
  \label{fig:cdfs}
\end{figure}

\begin{figure}[t]
  \centering
  \includegraphics{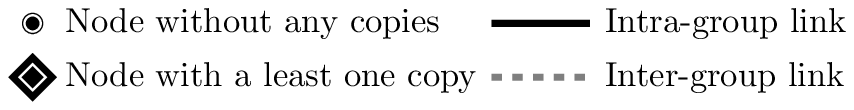} \\
  \subfloat[HYMAD: success within 15 seconds. \label{swg15}]{\includegraphics{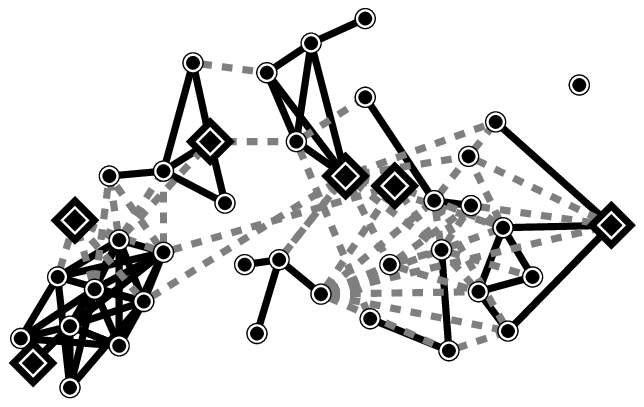}} \quad
  \subfloat[Spray-and-Wait: the copies stagnate around the source. It will take a total of 525 seconds to hit the destination. \label{sw15}]{\includegraphics{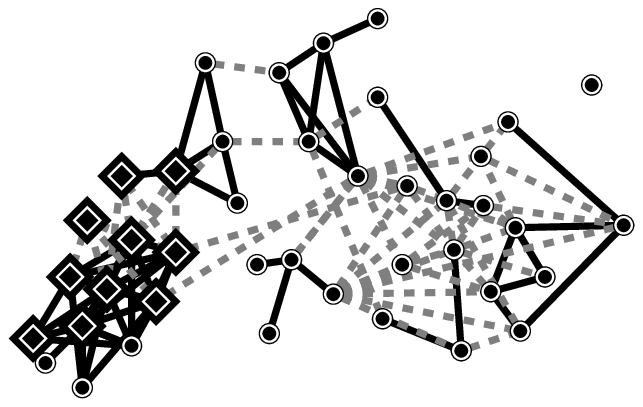}}
  \caption{Regular vs Hybrid Spray and Wait routing in Rollernet. (Partial view of the topology at t=15s)}
  \label{fig:diffusion}
\end{figure}

To evaluate the performance of HYMAD we replayed 3000 seconds of
the trace. This averages results over both the connected and
disconnected phases of Rollernet. Fig.~\ref{fig:cdfs} presents the
cumulative distribution function of the delivery probability for HYMAD
(with 5 copies), Spray-and-Wait (with 5 copies) and Epidemic. A few
observations can be made:
\begin{itemize}
\item HYMAD clearly outperforms Spray-and-Wait in terms of delay and
  quickly achieves comparable performance with Epidemic.
\item HYMAD also outperforms Spray-and-Wait in terms of delivery ratio
  for reasons explained hereafter.
\end{itemize}

Spray-and-Wait's simple forwarding scheme performs very well
\emph{under the assumption of independent and identically distributed
  node mobility}~\cite{spyro_sw}. However this is absolutely
\emph{not} the case in Rollernet where groups of friends tend to stick
together. It is also usually \emph{not} the case in many real-world
situations where underlying social dynamics are often at work.

This can have a impact on performance. For example, when using just 5
copies, Spray-and-Wait simply fails to deliver about 5\% of messages
even after waiting for more than 15 minutes. The
average delay with Spray-and-Wait ($210 \pm 14$ seconds) is nearly double
that of HYMAD ($120 \pm 6$ seconds). To further illustrate this point,
Fig~\ref{fig:diffusion} compares the propagation of 10 copies after 15
seconds for HYMAD (Fig.~\ref{swg15}) and Spray-and-wait
(Fig.~\ref{sw15}). The rightmost node is the head of the rollerblading
tour. The bold lines represent intra-group links while the dashed gray
lines represent inter-group links. The nodes holding at least one copy
are represented by a diamond. In HYMAD's case, the destination is a
diamond meaning that our hybrid approach has delivered its message
within 15 seconds. On the other hand, the regular Spray-and-Wait
protocol distributed copies mainly within its own local group. These
nodes remain close to each other thus increase the delay. In this
particular case (Fig.~\ref{sw15}) it will take 525 seconds for a node
with a copy to meet the destination

\section{Conclusion and further work}
\label{sec:conclusion}
In this paper we identified a new class of dense and highly mobile
networks not well addressed by conventional DTN or MANET approaches. We
proposed a new generic hybrid approach, HYMAD, that uses nodes' knowledge of
their local group topology to improve the performance of a simple DTN
protocol. In our case we used diameter-constrained groups along with
distance vector for intra-group routing and Spray-and-Wait for
inter-group routing. Simulations of our implementation in a dense and
highly mobile network show significant performance improvements over
regular Spray-and-Wait.

HYMAD is an example of a larger class of hybrid DTN-MANET routing
protocols which can handle a very wide spectrum of networks that
overlaps with those usually handled by either DTN or MANET.  We
believe that the first results that we obtained are encouraging for
further research in this direction. In particular, does $D_{max}$ need
to be a network-wide parameter? Could it be dynamic and would nodes
have to agree on it? More generally, different group partitioning
algorithms would be worth exploring, while other more elaborate
DTN/MANET protocol pairs could conceivably be used for intra and
inter-group routing.

\section*{Acknowledgments}
This work has been partially supported by the ANR project Crowd under
contract ANR-08-VERS-006.

\bibliographystyle{elsarticle-num}

\begin{thebibliography}{10}
\expandafter\ifx\csname url\endcsname\relax
  \def\url#1{\texttt{#1}}\fi
\expandafter\ifx\csname urlprefix\endcsname\relax\def\urlprefix{URL }\fi
\expandafter\ifx\csname href\endcsname\relax
  \def\href#1#2{#2} \def\path#1{#1}\fi

\bibitem{whitbeck-hymad}
J.~Whitbeck, V.~Conan, {HYMAD: Hybrid DTN-MANET Routing for Dense and Highly
  Dynamic Wireless Networks}, in: {AOC'09: Proceedings of the Third IEEE WoWMoM
  Workshop on Autonomic and Opportunistic Communications}, 2009.

\bibitem{ONE}
A.~Ker\"{a}nen, J.~Ott, T.~K\"{a}rkk\"{a}inen, {The ONE Simulator for DTN
  Protocol Evaluation}, in: SIMUTools '09: Proceedings of the 2nd International
  Conference on Simulation Tools and Techniques, 2009.

\bibitem{mcnamara}
L.~McNamara, C.~Mascolo, L.~Capra, Media sharing based on colocation prediction
  in urban transport, in: MobiCom '08: Proceedings of the 14th ACM
  international conference on Mobile computing and networking, 2008.

\bibitem{claveirole}
T.~Claveirole, M.~Boc, M.~D. de~Amorim, An empirical analysis of wi-fi activity
  in three urban scenarios, in: PERCOM '09: Proceedings of the 2009 IEEE
  International Conference on Pervasive Computing and Communications, 2009.

\bibitem{harri:models}
J.~Harri, F.~Filali, C.~Bonnet, Mobility models for vehicular ad hoc networks:
  a survey and taxonomy, IEEE Communications Surveys \& Tutorials 11~(4) (2009)
  19--41.

\bibitem{Burgess:2006}
J.~Burgess, B.~Gallagher, D.~Jensen, B.~N. Levine, {MaxProp: Routing for
  Vehicle-Based Disruption-Tolerant Networks}, in: Proc. {IEEE} Infocom, 2006.

\bibitem{crawdad}
{CRAWDAD}: A community resource for archiving wirelessdata at dartmouth,
  \url{http://crawdad.cs.dartmouth.edu}.

\bibitem{dtn_fall_sigcomm}
K.~Fall, A delay-tolerant network architecture for challenged internets, in:
  Proc. {ACM SIGCOMM}, 2003.

\bibitem{GrossglauserTse2002}
M.~Grossglauser, D.~N.~C. Tse, Mobility increases the capacity of ad hoc
  wireless networks, IEEE/ACM Trans. Netw. 10~(4) (2002) 477--486.

\bibitem{lindgren03}
A.~Lindgren, A.~Doria, O.~Schelen, Probabilistic routing in intermittently
  connected networks, in: Proc. {SAPIR}, 2004.

\bibitem{daly07}
E.~Daly, M.~Haahr, Social network analysis for routing in disconnected
  delay-tolerant {MANETs}, in: Proc. {ACM MobiHoc}, 2007.

\bibitem{LER}
M.~Grossglauser, M.~Vetterli, Locating nodes with ease: Last encounter routing
  in ad hoc networks through mobility diffusion, in: Proc. IEEE Infocom, 2003.

\bibitem{JOTT06}
J.~Ott, D.~Kutscher, C.~Dwertmann, Integrating dtn and manet routing, in:
  CHANTS '06: Proceedings of the 2006 SIGCOMM workshop on Challenged networks,
  2006.

\bibitem{r_operators}
S.~Dela\"et, B.~Ducourthial, S.~Tixeuil, Self-Stabilizing Systems, Springer,
  2005, Ch. Self-stabilization with r-Operators Revisited, pp. 68--80.

\bibitem{DKP08}
B.~Ducourthial, S.~Khalfallah, F.~Petit, Best effort group service in dynamic
  networks, Tech. rep. (2008).

\bibitem{spyro_sw}
T.~Spyropoulos, K.~Psounis, C.~Raghavendra, Spray and wait: an efficient
  routing scheme for intermittently connected mobile networks, in: WDTN '05:
  Proceedings of the 2005 ACM SIGCOMM workshop on Delay-tolerant networking,
  2005.

\bibitem{tournoux08_rollernet}
P.-U. Tournoux, J.~Leguay, F.~Benbadis, V.~Conan, M.~D. de~Amorim, J.~Whitbeck,
  The accordion phenomenon: Analysis, characterization, and impact on dtn
  routing, in: Proc. IEEE Infocom, 2009.

\bibitem{AMKA08}
D.~Antonellis, A.~Mansy, K.~Psounis, M.~H. Ammar, Towards distributed network
  classification for mobile ad hoc networks, in: WICON '08: Proceedings of the
  4th Annual International Conference on Wireless Internet, 2008.

\bibitem{BorrelAmmar07}
V.~Borrel, M.~H. Ammar, E.~W. Zegura., Understanding the wireless and mobile
  network space: a routing-centered classification, in: CHANTS '07: Proceedings
  of the second ACM workshop on Challenged networks, 2007.

\bibitem{ZhaoFerries}
W.~Zhao, M.~Ammar, E.~Zegura, A message ferrying approach for data delivery in
  sparse mobile ad hoc networks, in: MobiHoc '04: Proceedings of the 5th ACM
  international symposium on Mobile ad hoc networking and computing, 2004.

\bibitem{Shah2003}
R.~C. Shah, S.~Roy, S.~Jain, W.~Brunette, Data mules: Modeling a three-tier
  architecture for sparse sensor networks, in: IPSN 2008: Proceedings of the
  First IEEE Workshop on Sensor Network Protocols and Applications, 2003.

\bibitem{ClusteringSurvey}
J.~Yu, P.~Chong, A survey of clustering schemes for mobile ad hoc networks,
  {Communications Surveys \& Tutorials, IEEE} 7~(1) (2005) 32--48.

\bibitem{rollernet_youtube}
J.~Whitbeck, Animation of rollernet connectivity graph,
  \mbox{\url{http://www.youtube.com/watch?v=kdkCx1xlMkI}}.

\end{thebibliography}

\end{document}